\documentclass[11pt,a4paper]{article}
\usepackage[utf8]{inputenc}
\usepackage{amsmath}
\usepackage{amsfonts}
\usepackage{hyperref}
\usepackage{amssymb}
\usepackage{makeidx}
\usepackage{cite}
\usepackage{bm}
\usepackage{geometry}
\usepackage{doc}

\usepackage{url}

\usepackage{graphicx}
\usepackage{epstopdf}
\begin{document}

\begin{titlepage}

~~\\

\vspace*{0cm}
    \begin{Large}
    \begin{bf}
       \begin{center}
         {Wald type analysis for spin-one fields in three dimensions}
       \end{center}
    \end{bf}   
    \end{Large}
\vspace{0.7cm}
\begin{center} 
Nabarun Bera\footnote
            {
e-mail address : 
nabarunbera120892@gmail.com},
Suchetan Das\footnote
            {
e-mail address : 
suchetan.das@rkmvu.ac.in},
   
Bobby Ezhuthachan\footnote
            {
e-mail address : 
bobby.ezhuthachan@rkmvu.ac.in}\\

\vspace{0.7cm}
{\it Ramakrishna Mission Vivekananda University, Belur Math, Howrah-711202, West Bengal, India}\\

\end{center}

\vspace{0.7cm}

\begin{abstract}
We revisit Wald's analysis of \cite{Wald:1986bj} in the context of spin-one fields in three dimensions. A key technical difference from Wald's is the role played by the three dimensional completely antisymmetric tensor. We show how this changes the analysis as well as the result from that of \cite{Wald:1986bj} 
\end{abstract}
\end{titlepage}
\pagenumbering{arabic}
\tableofcontents

\section{Introduction}\label{intro}



  
The problem of finding interacting theories of spin-one and spin-two fields has a long history starting from the the works of \cite{Weinberg:1965rz},\cite{Feynman:2003hj}, \cite{Deser:1969wk}, \cite{Boulware:1975bs}. Since then, interactions for  both massive\cite{Hinterbichler:2011tt}, and massless spin-one and spin-two as well as for higher spin fields\cite{Vasiliev:1995dn} have been studied in a wide variety of theories using various methods: eg cohomological methods\cite{Henneaux:1997bm}, by imposing causality constraints\cite{Hertzberg:2017abn}, and demanding existance of a non-trivial S-Matrix \cite{Benincasa:2007xk},\cite{Benincasa:2011pg} among others\footnote{See references \cite{Deser:1987uk}-\cite{Hertzberg:2017abn} for a partial list of relevant papers in this direction.}.
One of the earliest works on such a classification of consistent spin-one and spin-two field theories was initiated by Wald  in \cite{Wald:1986bj}, where a systematic classical analysis had been undertaken to find all types of `consistent' self interactions of spin-one and spin-two fields, starting from the free theory. The `consistency condition' discussed therein results from the abelian gauge invariance of the free part of the Lagrangian which propagates to the higher order terms in the action and constrains the equations of motion of the full interacting theory. Solving these constraints give rise to non-abelian Yang-Mills type gauge invariance and general covariance for the spin-one and spin-two fields respectively\footnote {For the spin two field, the result of their analysis shows that apart from general covariance the consistency conditions are also satisfied by a class of theories with normal spin-two gauge invariance.}. 

In this note, we have tried to redo the analysis of Wald for spin-one fields in three dimensions with the key technical difference  being the role played by the three-dimensional totally antisymmetric tensor. Before discussing our results, we begin with a brief summary of Wald's analysis (for spin-one fields) as presented in \cite{Wald:1986bj}. Wald starts with a very general Lagrangian, the quadratic part of which is the sum of independent Maxwell terms, one for each gauge field $A^{\mu}_{a}$. Following Wald's notation, we have labelled the species-number of gauge fields by $\mu$ and the space-time index by $a$. This quadratic piece is denoted by $\mathcal{L}^{(2)}$ below. The higher order self-interaction terms, whose form we do not know is denoted by $\mathcal{L}^{(n)}$, where $n$ labels the order of the interaction. For example $n=3$ corresponds to cubic interactions, $n=4$ to quartic and so on. 
\begin{equation}
\mathcal{L} = \mathcal{L}^{(2)} + \sum^{\infty}_{n=3} \mathcal{L}^{(n)}
\end{equation}
The corresponding equation of motion for each species($F^{\mu}_a$) can again be expressed as a sum of terms, the first being linear obtained from the Maxwell part of the Lagrangian. These are denoted by $F^{(1)\mu}_{a}$ and $F^{(n)\mu}_{a}$ respectively. 
\begin{equation}
F^{\mu}_{a} = F^{(1)\mu}_{a} + \sum_{n=2}F^{(n)\mu}_{a}
\end{equation}

One then assumes that there exists a solution ($A^{\mu}_a$) to the full equation of motion(eom) $F^{\mu}_a$, which is perturbatively close to the solution of the free theory, which is taken to be $A^{\mu}_a =0$ for simplicity of analysis. This means that we assume that the equation of motion of the interacting theory has a solution of the type
\begin{equation}
A^{\mu}_a = \lambda\dot{A^{\mu}_{a}}+\lambda^{2}\ddot{A^{\mu}_{a}}+\dots
\end{equation}
where $\lambda$ is the perturbative parameter. Substituting this solution back into the eom and solving it order by order in $\lambda$, one gets equations to be satisfied by each of the $\dot{A}$, $\ddot{A}$, $\dots$\footnote{We are following Wald's notation wherein the $\dot{A}$, $\ddot{A}$, $\dots$ characterize the successive perturbative terms} etc. The equations arising from the two lowest orders in $\lambda$ are as follows: 
\begin{equation}
F^{(1)\mu}_{a}(\dot{A}) =0, \; \; F^{(2)\mu}_{a}(\dot{A}) + F^{(1)\mu}_a(\ddot{A}) =0
\end{equation}
The key point is that the Maxwell term $F^{(1)\mu}_{a}$, satisfies the divergence identity $\partial^{a} F^{(1)\mu}_{a}=0$. From the above equation, it is clear that this implies that $\dot{A}$ satisfies the further constraint $\partial^{a}F^{(2)\mu}_{a}(\dot{A})=0$. By following the same logic, we will end up with such constraints at higher orders. The nth order perturbation term thus has to satisfy these constraints, over and above the equation of motion. The origin of these constraints lies in the form of the quadratic piece, and therefore in the abelian gauge invariance satisfied by the Maxwell equation. Through the interaction terms, this propagates to the higher order terms. Wald's `consistency requirement' comes from demanding a single identity for the full interacting Lagrangian\footnote{Details will be given in the section \ref{single}}, which would result in the various constraints that  we see at various orders, in the perturbative analysis. This identity implies a set of infinitesimal gauge transformation under which the full action is invariant. Wald explicitly writes down an expression of the general form of such transformations $\delta_{\chi} A^{\mu}_{a} = \beta^{b\mu}_{a\nu}(\partial_{b}\chi^{\nu}+\alpha^{\nu}_{b\lambda}\chi^{\lambda})$ for some arbitrary function $\chi$. Wald constructs the functions $\beta$ and $\alpha$ out of the gauge fields, their derivatives as well as the invariant metric tensor $\eta^{ab}$. By imposing integrability conditions on such transformations, one can constrain the structure of the invariance further and show that they are precisely the non-abelian Yang-Mills type gauge transformations. The integrability condition is the condition that the commutator of two gauge transformations closes into a gauge transformation ${\it offshell}$, ie without imposing equations of motion:
\begin{equation}\label{int}
[\delta_{\phi}, \delta_{\psi}] A^{\mu}_{a} = \delta_{\chi}A^{\mu}_{a}
\end{equation}


In this note we redo Wald's analysis in three dimensions, with the only difference  that we demand that the functions $\beta$ and $\alpha$ that appear in the gauge transformations, are not just functions of $A^{\mu}$, its derivatives and the metric tensor $\eta^{ab}$ but 
also of the antisymmetric tensor $\epsilon^{abc}$. In three dimensions, we can start from either the Maxwell type quadratic action, or the Chern-Simons type quadratic action. Wald's analysis is not sensitive to the form of the quadratic piece, but only to its gauge-invariance.

Our analysis shows that introducing the $\epsilon$ term changes the analysis, and so at the infinitesimal level, unlike the usual gauge invariance \textbf{$\delta A^{\mu}_{a}=\partial_{a}\chi^{\mu}+f^{\mu}_{\lambda\sigma}(A^{\sigma}_{a}\chi^{\lambda})$}, we get a modified gauge invariance which involves arbitrary order of the gauge field. 
\begin{equation}
\delta A^{\mu}_{a} = \partial_{a}\chi^{\mu} + f^{\mu}_{\lambda\sigma}A^{\sigma}_{a}\chi^{\lambda} + d^{\nu\mu}_{\lambda}\epsilon_{abc}\partial^b A^{c}_{\nu}\chi^{\lambda} +\mathcal{O}(A^2)\chi+\mathcal{O}(A^2)\partial\hspace{.05 mm}\chi
\end{equation}
where $f^{\mu\nu}_{\lambda}$ is the structure constant of the gauge group.
One solution to the Wald-type analysis is  to put the coefficients of the $\epsilon$ terms, $d^{\mu\nu}_{\lambda}$ consistently to be zero. However as we discuss later, this is not the most general solution.  

A special case of our analysis is when there is only one species of gauge field. In this case, $f=0$ and $d$ is just one number. We find that even in this case there are non-trivial solutions corresponding to non-zero $d$'s. This special case has already been analyzed long back in the literature by Heiderich and Unruh \cite{Heiderich:1990bb}. Our answer reduces to that of \cite{Heiderich:1990bb} in this limit. 

\begin{itemize}

\item{{\bf clarification}:} In a previous version of the draft, we had made a mistake in analyzing the solutions for the single species case and claimed that the only possible solution is usual abelian gauge invariance. However after correcting the mistake, our solution matches with\cite{Heiderich:1990bb}. We had also missed this reference prevously. 
\end{itemize}

For the general case with multiple species of gauge fields, we get non trivial constraints which must be satisfied by $f$ and $d$ given in \ref{73}-\ref{75}, one of which is the well known Jacobi identity for the $f$'s.  To the best of our knowledge, this more general case has not been analyzed before. 

In the next section, we present the details of our analysis. In some examples, we find explicit solutions  
 for the equations \ref{73}-\ref{75}. In particular for the case of six species of gauge fields and $f$ corresponding to SO(4) algebra, we show that solutions exist with non zero $d$. Also, for $f=0$, the only non trivial equation that $d$'s have to satisfy is \ref{74}, for which non trivial solutions exist.
 
 Even though Walds analysis does seem to admit non linear gauge transformations as solutions to the integrability condition\ref{int}, one can ask whether there exists Lagrangians which are invariant under these new gauge transformations. Cohomological analysis that already exist in the literature, imply that the only gauge theories are of the Yang-Mills or Chern-Simons type in three dimensions\cite{Henneaux:1997bm}. We end with a discussion  on these issues in section \ref{lag}.

\section{Consistency analysis for interacting spin-one fields in three dimensions}  \label{single}
In this section,  we discuss the possible self-couplings of spin-one fields, following Wald's analysis. 
The general interacting Lagrangian is given as:
\begin{equation}
\mathcal{L} \equiv \mathcal{L}^{(2)}+ \sum_{n=3} \mathcal{L}^{(n)} 
\end{equation}
The quadratic piece is a sum of free Maxwell or Chern-Simons terms for each spin-one field\footnote{Instead of choosing sum of various terms in the quadratic piece, one may even alternate between sum and difference of terms. The analysis does not depend on this choice. In a more general setting one can also consider non diagonal quadratic kinetic terms which we discuss in a later section.}. The analysis does not depend on the precise form, but rather the gauge invariance of the quadratic part of the full action. 
We start with a collection of k spin one fields $A_{a}^{1},A_{a}^{2},\dots A_{a}^{k}$. We can label it in a single form $A_{a}^{\mu}$ where $\mu = 1,2, \dots k$. Here Greek letters label the species number.

The Lagrangian for a single free Chern Simons field in three dimension $\mathcal{L}_{cs}^{(2)}$ is of the form 
\begin{equation}
\mathcal{L}_{cs}^{(2)} = \frac{\kappa}{2}\epsilon^{abc}A_{a} \partial_{b} A_{c}
\end{equation}
The equation of motion is 
\begin{equation}
F^{(1)a}\equiv\frac{\kappa}{2}\epsilon^{abc}\mathcal{F}_{bc}=0
\end{equation}
Where $\mathcal{F}_{ab}=\partial_{a}A_{b}-\partial_{a}A_{b}$ is the field strength. 
The chern-simons field satisfies the divergence identity relation
\begin{equation}
\partial_{a} F^{(1)a} \equiv \frac{\kappa}{2}\epsilon^{abc} \partial_{a}\mathcal{F}_{bc} = 0
\end{equation}
Similar treatment can also be applied for the Maxwell case which was originally considered by Wald in four dimension. The Maxwell Lagrangian is
\begin{align*}
\mathcal{L}_{max}^{(2)} = \frac{1}{4}\mathcal{F}_{ab}\mathcal{F}^{ab}
\end{align*}  
with the corresponding eom being 
\begin{align*}
F^{(1)a}=\partial_{b}\mathcal{F}^{ba}=0
\end{align*}
which again satisfies the same divergence identity as the Chern-Simons case:
\begin{align*}
\partial_{a} F^{(1)a}=\partial_{a}\partial_{b}\mathcal{F}^{ba}=0
\end{align*}
So the  divergence identity $\partial_{a} F^{(1)a} = 0$ holds for both type of fields and therefore in what follows the same discussion applies for both cases.
Following Wald, we would like to find a consistent non-linear generalization for this free theory.

The equation of motion, for each species, of this non linear Lagrangian is 
\begin{equation}
F^{a}_{\mu}=\frac{\delta \mathcal{S}}{\delta A^{\mu}_{a}}
\end{equation}
Where $\mathcal{S}$ is the nonlinear action. Following Wald, we assume that a solution $A^{\mu}_{a}$ exists of the non-linear eom, which is arbitrarily close to the solution of the free theory, which we choose to be $A^{(free)\mu}_{a}=0$. Therefore, we take a solution of the form:
\begin{equation}
A^{\mu}_{a}=0+\lambda\dot{A^{\mu}_{a}}+\lambda^{2}\ddot{A^{\mu}_{a}}+\dots
\end{equation}
where as mentioned in the introduction, $\lambda$ is the perturbative expansion parameter and $\dot{A_{a}},\ddot{A_{a}},\dots$ are the higher order perturbation terms.
Substituting in the equation of motion:
\begin{equation}
F^{a}_{\mu}=F^{a(1)}_{\mu}(\lambda\dot{A_{a}}+\lambda^{2}\ddot{A_{a}}+\dots)+F^{a(2)}_{\mu}(\lambda\dot{A_{a}}+\lambda^{2}\ddot{A_{a}}+\dots)+\dots = 0
\end{equation}
Solving this order by order in $\lambda$ we get the following equations
\begin{equation}\label{linear}
F^{a(1)}_{\mu}(\dot{A_{a}})=0
\end{equation}
This implies that linearized mode of $A^{\mu}_{a}$ satisfies the linear order equation of motion.
At order $\lambda^{2}$ we get,
\begin{equation}
F^{a(1)}_{\mu}(\ddot{A_{a}})+F^{a(2)}_{\mu}(\dot{A_{a}})=0
\end{equation}
Using the divergence identity we get:
\begin{equation}\label{constraint-linear}
\partial_{a} F^{a(2)}_{\mu}(\dot{A_{a}}) = 0
\end{equation}
This implies that linearized perturbation $\dot{A^{\mu}_{a}}$ must satisfy the equation (\ref{constraint-linear}) in addition to its equation of motion (\ref{linear}). 

Its easy to see that this is true for higher order perturbations as well. For example at order $\lambda^{3}$, we have
\begin{equation}\label{eom order2}
F^{(1)a\mu}(\dddot{A_{a}})+F^{a(2)}_{\mu}(\dot{A_{a}},\ddot{A_{a}})+F^{a(3)}_{\mu}(\dot{A_{a}}) = 0
\end{equation}
and using the divergence identity, 
\begin{equation}\label{constraint-quadratic}
\partial_{a}F^{a(2)}_{\mu}(\dot{A_{a}},\ddot{A_{a}})+\partial_{a}F^{a(3)}_{\mu}(\dot{A_{a}}) = 0
\end{equation}
So, $\ddot{A^{\mu}_{a}}$  has to satisfy the equation (\ref{constraint-quadratic}) in addition to its eom (\ref{eom order2}). So, at every order, the $\dot{A^{\mu}_{a}}$,            
$\ddot{A^{\mu}_{a}}$ etc have to satisfy more than one equation. So in general, there may be no consistent solutions for these perturbations. This is the consistency problem discussed in\cite{Wald:1986bj}. Wald's solution to this problem is to demand the existence of an identity involving the full $F^{(a)}_{\mu}$, which in the perturbative analysis, would reduce to precisely the additional equations that we encountered. In a derivative expansion of 
$F^{(a)}_{\mu}$, the form of the identity is given as: 
\begin{equation}\label{walddivergence}
\partial_{a}F^{a}_{\mu} = \lambda^{\nu}_{a\mu}F^{a}_{\nu} + \rho^{b\nu}_{a\mu}\partial_{b}F^{a}_{\nu}+\sigma^{bc\nu}_{a\mu}\partial_{b}\partial_{c}F^{a}_{\nu}
+\dots
\end{equation}
The ${\lambda^{\mu}_{a\nu},\rho^{b\mu}_{a\nu},\sigma^{bc\mu}_{a\nu},\dots}$ are made locally out of $\eta_{ab}$, $\epsilon^{abc}$, $A^{\mu}_{a}$ and derivatives of $A^{\mu}_{a}$'s. This identity must reduce to the divergence identity satisfied by the free Lagrangian, which implies that the functions ${\lambda^{\mu}_{a\nu},\rho^{b\mu}_{a\nu},\sigma^{bc\mu}_{a\nu},\dots}$ vanish when $A= A^{free}$.
Expanding (\ref{walddivergence}) perturbatively we get,
\begin{align}
\partial_{a}[F^{a(1)}_{\mu}(\lambda\dot{A}+\lambda^{2}\ddot{A}+\dots)+F^{a(2)}_{\mu}\nonumber
(\lambda\dot{A}+\lambda^{2}\ddot{A}+\dots)+\dots]= \\ \nonumber
\lambda_{a\nu}^{\mu(1)}[F^{a(1)}_{\nu}(\lambda\dot{A}+\lambda^{2}\ddot{A}+\dots)+F^{a(2)}_{\nu}(\lambda\dot{A}+\lambda^{2}\ddot{A}+\dots)+\dots]+ \\ \nonumber
\rho^{b\nu(1)}_{a\mu}\partial_{b}[F^{a(1)}_{\nu}(\lambda\dot{A}+ 
\lambda^{2}\ddot{A}+\dots)+F^{a(2)}_{\nu}(\lambda\dot{A}+\lambda^{2}\ddot{A}+\dots)+\dots]+ \\ \nonumber
\sigma_{a\mu}^{bc\nu(1)}\partial_{b}\partial_{c}[F^{a(1)}_{\nu}(\lambda\dot{A}+\lambda^{2}\ddot{A}+  \dots)+F^{a(2)}_{\nu}(\lambda\dot{A}+\lambda^{2}\ddot{A}+\dots)+\dots]+\dots \\
\end{align}
At order $\lambda$ this reduces to,
\begin{align}
\partial_{a}F^{a(1)}_{\mu}(\dot{A}) = 0
\end{align}
Which is linearized divergence identity as desired. At order $\lambda^{2}$ we get,
\begin{equation}
\partial_{a}F^{a(1)}_{\mu}(\ddot{A})+\partial_{a}F^{a(2)}_{\mu}(\dot{A}) = \lambda_{a\mu}^{(1)\nu}F^{a(1)}_{\nu}(\dot{A})+\rho_{a\mu}^{b\nu(1)}\partial_{a}F^{a(1)}_{\nu}(\dot{A})+\sigma_{a\mu}^{bc\nu(1)}\partial_{b}\partial_{c}F^{a(1)}_{\nu}(\dot{A})+\dots
\end{equation} 
Using the linearized identity relation as well as as the linearized equation of motion $F^{(1)a\mu}(\dot{A})=0$ we get,
\begin{equation}
\partial_{a}F^{a(2)}_{\mu}(\dot{A})=0
\end{equation}
Thus the identity (\ref{walddivergence}) and the linearized eom $F^{a(1)}_{\mu}(\dot{A})=0$ implies $\partial_{a}F^{a(2)}_{\mu}(\dot{A_{a}})=0$. In this way, order by order, all the additional equations, are automatically satisfied, by demanding the identity (\ref{walddivergence}) as well as the eom at that order.

Following Wald, we further assume that the derivative expansion in (\ref{walddivergence}) truncates at first order, and also that the coefficients  $\lambda^{\mu}_{a\nu}$ contains no more than one derivative of $A^{\mu}_{a}$ while $\rho^{b\mu}_{a\nu}$ contains no derivatives of $A^{\mu}_{a}$. As mentioned in \cite{Wald:1986bj}, this is one of the strongest assumption of the entire analysis. This is a simplifying assumption which ensures that at any perturbative order, the number of derivatives on both side of (\ref{walddivergence}) are the same. Then (\ref{walddivergence}) takes the form, 
\begin{align}
\partial_{a}F^{a\mu}=\lambda^{\mu}_{a\nu}F^{a\nu} + \rho^{b\mu}_{a\nu}\partial_{b}F^{a\nu}
\end{align}
 
which after a redefinition, can be expressed as: 

\begin{align}\label{walddivergencesimple}
\partial_{b}(\beta^{b\nu}_{a\mu}F^{a}_{\nu})=\beta^{b\nu}_{a\lambda}\alpha^{\lambda}_{b\mu}F^{a}_{\nu}
\end{align}

where $\beta_{a\nu}^{b\mu} = \delta_{a}^{b}\delta_{\nu}^{\mu} - \rho_{a\nu}^{b\mu}$ and $\lambda^{\mu}_{a\nu}=\beta_{a\lambda}^{b\mu}\alpha^{\lambda}_{b\nu}-\partial_{b}\rho^{b\mu}_{a\nu}$.

It is easy to see that under the following transformation, (\ref{walddivergencesimple}) remains unchanged. 
\begin{eqnarray}\label{1}
&&\beta^{b\mu}_{a\nu} \rightarrow f^{\lambda}_{\nu}\beta^{b\mu}_{a\lambda} \nonumber \\
&&\alpha^{\mu}_{a\nu} \rightarrow (f^{-1})^{\mu}_{\sigma}f^{\delta}_{\nu}\alpha^{\sigma}_{a\delta} + (f^{-1})^{\mu}_{\kappa}\partial_{a}f^{\kappa}_{\nu}
\end{eqnarray}
Where $f$ is an arbitrary function locally made out of $A_{a}$ and equals 1 when $A_a=0$ . 
 
multiplying (\ref{walddivergencesimple}) by an arbitrary function $\chi(x)^{\mu}$ and integrating over space-time, 
\begin{align}
\nonumber
0=\int{d^3 x}\chi^{\nu}[\partial_b(\beta_{a\nu}^{b\mu} F ^a_{\mu})-\beta_{a\lambda}^{b\mu} \alpha^{\lambda}_{b\nu}F^a_{\mu}] \\ \nonumber
= \int{d^3 x}[\partial_b(\chi^{\nu}\beta_{a\nu}^{b\mu} F ^a_{\mu})-\beta_{a\nu}^{b\mu}F^{a}_{\mu}(\partial_{b}\chi^{\nu})-\chi^{\nu}\beta^{b\mu}_{a\lambda} \alpha^{\lambda}_{b\nu}F^a_{\mu}] \\ \nonumber
=-\int{d^3 x}\beta_{a\nu}^{b\mu}(\partial_b\chi^{\nu}+\alpha^{\nu}_{b\lambda} \chi^{\lambda})\frac{\delta \mathcal{S}}{\delta A^{\mu}_a}
\end{align}
Where in the second line we have used integration by parts and dropped the total derivative term.
Therefore, the identity (\ref{walddivergencesimple}) can be interpreted as demanding the invariance of the full action under the infinitesimal variation of the vector field\footnote{Note that here we are considering transformations which make the action invariant, not the Lagrangian.
Since we are not considering large gauge transformations, this includes the case of Chern Simons theory.}. 
\begin{align}\label{infinitesimalgauge}
\delta A^{\mu}_{a} = \beta^{b\mu}_{a\nu}(\partial_{b}\chi^{\nu}+\alpha^{\nu}_{b\lambda}\chi^{\lambda})
\end{align}

Demanding that this infinitesimal gauge transformation corresponds to a finite gauge symmetry of the action, imposes an integrability condition on the infinitesimal gauge transformation. In the language of differential geometry, $\delta_{\chi}A$ is the tangent vector that generates the transformation on the manifold formed by set of $A_a$'s. For some arbitrary function $\phi$ and $\psi$ on spacetime, the subspace of infinitesimal gauge transformations is generated by vector fields $V_{\phi}$ and $V_{\psi}$ respectively. The integrability condition is then determined by the Frobenius Theorem which states that there must exist a function $\chi$ such that the commutator $[V_{\phi},V_{\psi}]$ is equal to $V_{\chi}$. This means if the action is invariant under the transformation generated by $V_{\phi}$ and $V_{\psi}$, it must be invariant under the infinitesimal transformation generated by the commutator $[V_{\phi},V_{\psi}]$\footnote{More details are given in \cite{Wald:1986bj}}.
 
Thus we get, $[\delta_{\phi},\delta_{\psi}]A_{a} = \delta_{\chi}A_{a}$

\begin{align}\label{frobenius}
\implies \delta_{\phi}(\beta^{b\mu}_{a\nu})[\partial_{b}\psi^{\nu} + \alpha_{b\lambda}^{\nu}\psi^{\lambda}] - \delta_{\psi}(\beta^{b\mu}_{a\nu})[\partial_{b}\phi^{\nu} + \alpha_{b\lambda}^{\nu}\phi^{\lambda}]+\nonumber \\ 
\beta^{b\mu}_{a\nu}(\delta_{\phi}\alpha_{b\lambda}^{\nu})\psi^{\lambda} - \beta^{b\mu}_{a\nu}(\delta_{\psi}\alpha_{b\lambda}^{\nu})\phi^{\lambda} =\beta_{a\nu}^{b\mu}(\partial_b\chi^{\nu}+\alpha_{b\lambda}^{\nu}\chi^{\lambda})
\end{align}
Using (\ref{infinitesimalgauge}), we can express 
\begin{equation}
\delta_{\phi}\beta^{b\mu}_{a\nu} = \frac{\partial \beta^{b\mu}_{a\nu}}{\partial A_{c}^{\lambda}} \delta_{\phi}A_{c}^{\lambda} = \frac{\partial \beta^{b\mu}_{a\nu}}{\partial A_{c}^{\lambda}}\beta_{c\rho}^{d\lambda}(\partial_d\phi^{\rho}+\alpha_{d\sigma}^{\rho}\phi^{\sigma})
\end{equation}

So that the integrability condition becomes,
\begin{align}\label{multiple}
[\frac{\partial \beta^{b\nu}_{a\mu}}{\partial A_{c}^{\sigma}}\beta^{d\sigma}_{c\lambda} - \frac{\partial \beta^{d\nu}_{a\lambda}}{\partial A_{c}^{\sigma}}\beta^{b\sigma}_{c\mu}](\partial_{d}\phi^{\lambda} + \alpha_{d\rho}^{\lambda}\phi^{\rho})(\partial_{b}\psi^{\mu} + \alpha_{b\rho}^{\mu}\psi^{\rho})+\nonumber \\ 
\beta^{b\nu}_{a\sigma}(\delta_{\phi}(\alpha_{b\mu}^{\sigma})\psi^{\mu} - \delta_{\psi}(\alpha_{b\mu}^{\sigma})\phi^{\mu}) =\beta_{a\mu}^{b\nu}(\partial_b\chi^{\mu}+\alpha_{b\rho}^{\mu}\chi^{\rho})
\end{align}

So the main goal is to solve equation(\ref{multiple}) and get the expressions for $\beta_{a}^{b}$ and $\alpha_{b}$. To solve this we will follow the procedure as described by Wald. 

Expanding $\beta_{a}^{b}$, $\alpha_{b}$ and $\chi$ in power series of $A_{a}$, solve the equation order by order. i.e
\begin{equation*}
\alpha_{a} = \sum_{n}\alpha_{a}^{(n)}
\end{equation*}
\begin{equation*}
\beta_{a} = \sum_{n}\beta_{a}^{b(n)}
\end{equation*}
\begin{equation*}
\chi = \sum_{n}\chi^{(n)}
\end{equation*}
To solve the zeroth order part of the equation (\ref{multiple}) we need the expression for $\alpha^{(1)}_{a}$, $\beta_{a}^{b(1)}$, $\alpha^{(0)}_{a}$ and $\beta_{a}^{b(0)}$. For $A_{a}=0$ we have, that 
\begin{align}
\beta^{(0)b\nu}_{a\mu}=\delta^{b}_{a}\delta^{\nu}_{\mu}, \; \;  \;
\alpha^{(0)\lambda}_{b\mu}=0
\end{align}
Now $\alpha_{a}^{(1)}$ is an one index tensor that can be constructed locally from $\eta_{ab}$, $\epsilon_{abc}$ and $A_{a}$ or $\partial_{a}A_{b}$. Similarly $\beta_{a}^{b(1)}$ is the two index tensor that can be made by $\eta_{ab}$, $\epsilon_{abc}$ and $A_{a}$. So the general forms of $\alpha^{(1)}$ and $\beta^{(1)}$ are:
\begin{align}
\alpha_{a\mu}^{\nu(1)}= f^{\nu}_{\mu\lambda}A^{\lambda}_{a} + d^{\lambda\nu}_{\mu}\epsilon_{abc}\partial^{b}A^{c}_{\lambda} \\
\beta^{(1)b\nu}_{a\mu} = e^{\nu}_{\rho\mu}\epsilon^{bc}_{a}A^{\rho}_{c}
\end{align}

\textit{In this case $d^{\lambda\nu}_{\mu}$ and $e^{\nu}_{\rho\mu}$ are the new coefficients, that appear as a result of invoking the $\epsilon_{abc}$ tensor, change the form of $\alpha_{a\mu}^{\nu(1)}$ and $\beta^{(1)b\nu}_{a\mu} $ from that of \cite{Wald:1986bj}.}

The zeroth order part of equation (\ref{multiple}) gives,
\begin{align} \label{badeq}
[\frac{\partial \beta^{(1)b\nu}_{a\mu}}{\partial A_{c}^{\sigma}}\beta^{(0)d\sigma}_{c\lambda} - \frac{\partial \beta^{(1)d\nu}_{a\lambda}}{\partial A_{c}^{\sigma}}\beta^{(0)b\sigma}_{c\mu}](\partial_{d}\phi^{\lambda} + \alpha_{d\rho}^{(0)\lambda}\phi^{\rho})(\partial_{b}\psi^{\mu} + \alpha_{b\rho}^{(0)\mu}\psi^{\rho}) +\nonumber \\
\beta^{(0)b\nu}_{a\sigma}(\delta_{\phi}^{0}(\alpha_{b\mu}^{(1)\sigma})\psi^{\mu} - \delta_{\psi}^{(0)}(\alpha_{b\mu}^{(0)\sigma})\phi^{\mu})
=\beta_{a\mu}^{(0)b\nu}(\partial_b\chi^{(0)\mu}+\alpha_{b\rho}^{(0)\mu}\chi^{(0)\rho}) \nonumber \\
\implies  [e^{\nu}_{\lambda\mu}\epsilon^{bd}_{a} - e^{\nu}_{\mu\lambda}\epsilon^{db}_{a}]\partial_{d}\phi^{\lambda}\partial_{b}\psi^{\mu} +[\psi^{\mu}f^{\nu}_{\mu\lambda}\partial_{a}\phi^{\lambda} - \phi^{\mu}f^{\nu}_{\mu\lambda}\partial_{a}\psi^{\lambda}] + \nonumber \\ 
[\psi^{\mu}\frac{\partial \alpha^{(1)\nu}_{a\mu}}{\partial(\partial_{d}A^{\lambda}_{e})}\partial_{d}\partial{e}\phi^{\lambda} - \phi^{\mu}\frac{\partial \alpha^{(1)\nu}_{a\mu}}{\partial(\partial_{d}A^{\lambda}_{e})}\partial_{d}\partial{e}\psi^{\lambda}]\nonumber \\
= \partial_{a}\chi^{(0)\nu}
\end{align}
The term $\frac{\partial \alpha^{(1)\nu}_{a\mu}}{\partial(\partial_{d}A^{\lambda}_{e})}\partial_{d}\partial{e}\phi^{\lambda}$ vanishes due to antisymmetry property of $\epsilon_{abc}$.Thus the equation (\ref{badeq}) becomes,
\begin{align}\label{55}
[e^{\nu}_{\lambda\mu}\epsilon^{bd}_{a} + e^{\nu}_{\mu\lambda}\epsilon^{bd}_{a}]\partial_{d}\phi^{\lambda}\partial_{b}\psi^{\mu} +[\psi^{\mu}f^{\nu}_{\mu\lambda}\partial_{a}\phi^{\lambda} - \phi^{\mu}f^{\nu}_{\mu\lambda}\partial_{a}\psi^{\lambda}] = \partial_{a}\chi^{(0)\nu}
\end{align}
Taking curl on both side will give,
\begin{align}\label{e}
(e^{\nu}_{\lambda\mu}+e^{\nu}_{\mu\lambda})\epsilon^{bd}_{[a}\partial_{e]}(\partial_{d}\phi^{\lambda}\partial_{b}\psi^{\mu}) + f^{\nu}_{\mu\lambda}\partial_{[e}\psi^{(\mu}\partial_{a]}\phi^{\lambda )} = 0
\end{align}
Here we choose $\psi^{\mu} = x^{\mu}$ and $\phi^{\lambda} = x^{\lambda}$. This implies,
\begin{align}
f^{\nu}_{\mu\lambda}\delta_{[e}^{(\mu}\delta_{a]}^{\lambda )} = 0 \nonumber \\
\implies f^{\nu}_{\mu\lambda} = - f^{\nu}_{\lambda\mu}
\end{align}
So,we get $f^{\nu}_{\mu\lambda}$ is antisymmetric in $\mu$ and $\lambda$. Using this property the second term of (\ref{e}) is identically zero. So the equation (\ref{e}) implies,
\begin{align}
e^{\nu}_{\lambda\mu}\epsilon^{bd}_{[a}\partial_{e]}(\partial_{d}\phi^{\lambda}\partial_{b}\psi^{\mu}) + e^{\nu}_{\lambda\mu}\epsilon^{bd}_{[a}\partial_{e]}(\partial_{d}\phi^{\mu}\partial_{b}\psi^{\lambda}) = 0 \nonumber \\
\implies e^{\nu}_{\lambda\mu}\epsilon^{bd}_{[a}\partial_{e]}(\partial_{d}\phi^{\lambda}\partial_{b}\psi^{\mu} + \partial_{d}\phi^{\mu}\partial_{b}\psi^{\lambda}) = 0
\end{align}
This is true for all possible $\psi^{\mu}$ and $\phi^{\lambda}$. Thus we should get,
\begin{align}
e^{\nu}_{\lambda\mu} = 0
\end{align}
The antisymmetry property of $f^{\nu}_{\mu\lambda}$ and vanishing $e^{\nu}_{\lambda\mu}$ modify (\ref{55}) as following,
\begin{align}\label{1000}
\partial_{a}\chi^{(0)\nu} = \psi^{\mu}f^{\nu}_{\mu\lambda}\partial_{a}\phi^{\lambda} - \partial_{a}(\phi^{\mu}f^{\nu}_{\mu\lambda}\psi^{\lambda}) + \psi^{\lambda}f^{\nu}_{\mu\lambda}\partial_{a}\phi^{\mu} \nonumber \\
=- \partial_{a}(\phi^{\mu}f^{\nu}_{\mu\lambda}\psi^{\lambda}) + \psi^{\mu}f^{\nu}_{\mu\lambda}\partial_{a}\phi^{\lambda} - \psi^{\mu}f^{\nu}_{\mu\lambda}\partial_{a}\phi^{\lambda} \nonumber \\
= \partial_{a}(\psi^{\mu}f^{\nu}_{\mu\lambda}\phi^{\lambda}) \nonumber \\
\implies \chi^{(0)\nu} = f^{\nu}_{\mu\lambda}\psi^{\mu}\phi^{\lambda} + b^{\nu}
\end{align}
Where $b^{\nu}$ is a constant. So now we have $\beta^{(1)b\nu}_{a\mu} = 0$. But the constant term $d^{\lambda\nu}_{\mu}$, in $\alpha^{\nu(1)}_{a\mu}$ is left undetermined. So to get the value of this undetermined constant we have to study the first order part of (\ref{multiple}) and we will substitute the solution of $\beta^{(1)b\nu}_{a\mu}$ , $\alpha^{\nu(1)}_{a\mu}$ and $\chi^{(0)\nu}$ in there. Now We obtain,
\begin{align}\label{61}
[\frac{\partial \beta^{(2)b\nu}_{a\mu}}{\partial A_{c}^{\sigma}}\delta^{d}_{c}\delta^{\sigma}_{\lambda} - \frac{\partial \beta^{(2)d\nu}_{a\lambda}}{\partial A_{c}^{\sigma}}\delta^{b}_{c}\delta^{\sigma}_{\mu}]\partial_{d}\phi^{\lambda}\partial_{b}\psi^{\mu} +\delta^{b}_{a}\delta^{\nu}_{\sigma}[\psi^{\mu}\delta_{\phi}^{(0)}\alpha^{\sigma (2)}_{b\mu}-\phi^{\mu}\delta_{\psi}^{(0)}\alpha^{\sigma (2)}_{b\mu}\nonumber \\
+ \psi^{\mu}\frac{\partial \alpha^{\sigma (1)}_{b\mu}}{\partial A_{c}^{\lambda}}\delta^{d}_{c}\delta^{\lambda}_{\epsilon}(\alpha^{\epsilon (1)}_{d\zeta}\phi^{\zeta})-\phi^{\mu}\frac{\partial \alpha^{\sigma (1)}_{b\mu}}{\partial A_{c}^{\lambda}}\delta^{d}_{c}\delta^{\lambda}_{\epsilon}(\alpha^{\epsilon (1)}_{d\zeta}\psi^{\zeta}) \nonumber \\
+ \psi^{\mu}\frac{\partial \alpha^{\sigma (1)}_{b\mu}}{\partial(\partial_{e} A_{c}^{\lambda})}\delta^{d}_{c}\delta^{\lambda}_{\epsilon}\partial_{e}(\alpha^{\epsilon (1)}_{d\zeta}\phi^{\zeta})-\phi^{\mu}\frac{\partial \alpha^{\sigma (1)}_{b\mu}}{\partial(\partial_{e}A_{c}^{\lambda})}\delta^{d}_{c}\delta^{\lambda}_{\epsilon}\partial_{e}(\alpha^{\epsilon (1)}_{d\zeta}\psi^{\zeta})] \nonumber \\
=\delta^{b}_{a}\delta^{\nu}_{\mu}(\partial_{b}\chi^{(1)\mu} + \alpha^{\mu(1)}_{b\rho}\chi^{(0)\rho}) \nonumber \\
\implies [\frac{\partial \beta^{(2)b\nu}_{a\mu}}{\partial A_{d}^{\lambda}} - \frac{\partial \beta^{(2)d\nu}_{a\lambda}}{\partial A_{b}^{\mu}}]\partial_{d}\phi^{\lambda}\partial_{b}\psi^{\mu} + \psi^{\mu}\delta_{\phi}^{(0)}\alpha^{\nu (2)}_{a\mu} - \phi^{\mu}\delta_{\psi}^{(0)}\alpha^{\nu (2)}_{a\mu} \nonumber \\
+ [2f^{\nu}_{\mu \epsilon}f^{\epsilon}_{\zeta \delta}A^{\delta}_{a}+2(f^{\nu}_{\mu \epsilon}d^{\sigma \epsilon}_{\zeta}+f^{\sigma}_{\zeta \epsilon}d^{\epsilon \nu}_{\mu})\epsilon_{abc}\partial^{b}A^{c}_{\sigma}+2d^{\nu}_{\mu \epsilon}d^{\sigma \epsilon}_{\zeta}\epsilon^{d}_{ae}\epsilon_{dgh}\partial^{e}\partial^{g}A^{h}_{\sigma}]\phi^{[\zeta}\psi^{\mu ]} \nonumber \\
+(\psi^{\mu}\partial^{b}\phi^{\zeta}-\phi^{\mu}\partial^{b}\psi^{\zeta})[f^{\sigma}_{\zeta \epsilon}d^{\epsilon \nu}_{\mu}\epsilon_{abc}A^{c}_{\sigma}+d^{\nu}_{\mu \epsilon}d^{\sigma \epsilon}_{\zeta}\epsilon_{abc}\epsilon^{c}_{de}\partial^{d}A^{e}_{\sigma}] \nonumber \\
 = \partial_{a}\chi^{(1)\nu} + (f^{\nu}_{\rho\epsilon}A^{\epsilon}_{a}+d^{\sigma \nu}_{\rho}\epsilon_{abc}\partial^{b}A^{c}_{\sigma})[f^{\rho}_{\mu\zeta}\psi^{\mu}\phi^{\zeta}+b^{\rho}]
\end{align}\footnote{ The $d^{\nu}_{\mu\lambda}$ are defined in terms of the $d^{\lambda\nu}_{\mu}$ which appear in the gauge transformations as $d^{\nu}_{\mu\lambda}\equiv d^{\lambda\nu}_{\mu}$. We will follow this notation in the rest of the paper}
If we now take curl of the equation (\ref{61}), we see that left hand side is completely $\phi^\mu$ and $\psi^\mu$ dependent, whereas the part containing $b^{\nu}$ in the right hand side is not. As the equation should hold for all possible $\phi^\mu$ and $\psi^\mu$, $b^{\nu}$ must be zero. Now setting $\phi^\mu$ and $\psi^\mu$ to be constants we get, after taking curl 
\begin{align}\label{72}
2[f^{\nu}_{\mu \epsilon}f^{\epsilon}_{\zeta \sigma}\partial_{[e}A^{\sigma}_{a]}+(f^{\nu}_{\mu \epsilon}d^{\sigma \epsilon}_{\zeta}+f^{\sigma}_{\zeta \epsilon}d^{\epsilon \nu}_{\mu})\epsilon_{bc[a}\partial_{e]}\partial^{b}A^{c}_{\sigma}+d^{\nu}_{\mu \epsilon}d^{\sigma \epsilon}_{\zeta}\epsilon_{fd[a}\partial_{e]}\epsilon^{d}_{gh}\partial^{f}\partial^{g}A^{h}_{\sigma}]\phi^{[\zeta}\psi^{\mu ]} \nonumber \\
= (f^{\nu}_{\epsilon\sigma}\partial_{[e}A^{\sigma}_{a]}+d^{\sigma \nu}_{\epsilon}\epsilon_{bc[a}\partial_{e]}\partial^{b}A^{c}_{\sigma})f^{\epsilon}_{\mu\zeta}\psi^{\mu}\phi^{\zeta}
\end{align}
By simplifying equation (\ref{72}) with the help of antisymmetric property of $f^{\nu}_{\mu \epsilon}$, we will get the following equation.
\begin{align*}
[(f^{\nu}_{\mu \epsilon}f^{\epsilon}_{\zeta \sigma}-f^{\nu}_{\zeta \epsilon}f^{\epsilon}_{\mu \sigma}-f^{\nu}_{\epsilon \sigma}f^{\epsilon}_{\mu \zeta})\partial_{[e}A^{\sigma}_{a]} +(d^{\nu}_{\mu \epsilon}d^{\sigma \epsilon}_{\zeta}-d^{\nu}_{\zeta \epsilon}d^{\sigma \epsilon}_{\mu})\epsilon_{fd[a}\partial_{e]}\epsilon^{d}_{gh}\partial^{f}\partial^{g}A^{h}_{\sigma}+ \\
(f^{\nu}_{\mu \epsilon}d^{\sigma \epsilon}_{\zeta}-f^{\nu}_{\zeta \epsilon}d^{\sigma \epsilon}_{\mu}-f^{\epsilon}_{\mu \zeta}d^{\sigma \nu}_{\epsilon}+f^{\sigma}_{\zeta \epsilon}d^{\epsilon \nu}_{\mu}-f^{\sigma}_{\mu \epsilon}d^{\epsilon \nu}_{\zeta})\epsilon_{bc[a}\partial_{e]}\partial^{b}A^{c}_{\sigma})]\phi^{\zeta}\psi^{\mu} = 0
\end{align*}
This is the linearly independent equation where $\phi^\mu$ and $\psi^\mu$ are constants. Therefore we should have,
\begin{align}\label{73}
f^{\nu}_{\mu \epsilon}f^{\epsilon}_{\zeta \sigma}-f^{\nu}_{\zeta \epsilon}f^{\epsilon}_{\mu \sigma}-f^{\nu}_{\epsilon \sigma}f^{\epsilon}_{\mu \zeta} = 0 
\end{align}
\begin{align}\label{74}
d^{\nu}_{\mu \epsilon}d^{\sigma \epsilon}_{\zeta}-d^{\nu}_{\zeta \epsilon}d^{\sigma \epsilon}_{\mu} = 0 
\end{align}
\begin{align}\label{75}
f^{\nu}_{\mu \epsilon}d^{\sigma \epsilon}_{\zeta}-f^{\nu}_{\zeta \epsilon}d^{\sigma \epsilon}_{\mu}-f^{\epsilon}_{\mu \zeta}d^{\sigma \nu}_{\epsilon}+f^{\sigma}_{\zeta \epsilon}d^{\epsilon \nu}_{\mu}-f^{\sigma}_{\mu \epsilon}d^{\epsilon \nu}_{\zeta} = 0
\end{align}

The identity (\ref{73}) is precisely the Jacobi identity. So the antisymmetric property of $f^{\epsilon}_{\zeta\sigma}$ and the Jacobi identity suggests that it must be the structure constant of Lie group. However, there is a further parameter $d$, which satisfies the above two equations. Of course, one consistent solution is to choose $d=0$, in which case our analysis reduces to that of Wald's analysis, and we end up with the non-abelian lie algebra valued gauge invariance, as we discuss below. 
\begin{itemize}
\item For $d=0$, using (\ref{73}), equation (\ref{61}) becomes 
\begin{align}
[\frac{\partial \beta^{(2)b\nu}_{a\mu}}{\partial A_{d}^{\lambda}} - \frac{\partial \beta^{(2)d\nu}_{a\lambda}}{\partial A_{b}^{\mu}}]\partial_{d}\phi^{\lambda}\partial_{b}\psi^{\mu} + \psi^{\mu}\delta_{\phi}^{(0)}\alpha^{\nu (2)}_{a\mu} - \phi^{\mu}\delta_{\psi}^{(0)}\alpha^{\nu (2)}_{a\mu} = \partial_{a}\chi^{(1)\nu}
\end{align}
By using the procedure as discussed by Wald, of choosing trial functional forms of $\phi^{\mu}$ and $\psi^{\mu}$ we can show that $\delta_{\phi}^{(0)}\partial_{[e}\alpha^{\nu (2)}_{a]\mu} = 0$. Therefore we can set $\alpha^{\nu (2)}_{a\mu}$ to zero by the analogue of (\ref{1}). Thus for $n>2$ we have always $\alpha^{\nu (n)}_{a\mu} = 0$ and  $\beta^{(n)b\nu}_{a\mu} = 0$ by the method of induction used in \cite{Wald:1986bj} . So our solution for $\alpha^{\nu}_{a\mu}$ and $\beta^{b\nu}_{a\mu}$ in this case, is
\begin{align}
\beta^{b\nu}_{a\mu}=\delta^{b}_{a}\delta^{\nu}_{\mu}\\
\alpha^{\nu}_{a\mu}=f^{\nu}_{\mu\lambda}A^{\lambda}_{a} 
\end{align}
The gauge transformation corresponds to,
\begin{align}\label{81}
\delta A^{\mu}_{a}=\partial_{a}\chi^{\mu}+f^{\mu}_{\lambda\sigma}(A^{\sigma}_{a}\chi^{\lambda})\nonumber \\
=\partial_{a}\chi^{\mu}+[A_{a},\chi]^{\mu} 
\end{align}
Where $[,]$ is the Lie-algebra bracket.
\end{itemize}
 

\textit{However, this is not the general solution. There may exist solutions with  $d^{\sigma\epsilon}_{\zeta}\neq 0$.  This is the main difference from the result of \cite{Wald:1986bj}}. 
\begin{itemize}
\item non-zero $d$, equation (\ref{61}) becomes,
\begin{align}\label{82}
[\frac{\partial \beta^{(2)b\nu}_{a\mu}}{\partial A_{d}^{\lambda}} - \frac{\partial \beta^{(2)d\nu}_{a\lambda}}{\partial A_{b}^{\mu}}]\partial_{d}\phi^{\lambda}\partial_{b}\psi^{\mu} + \psi^{\mu}\delta_{\phi}^{(0)}\alpha^{\nu (2)}_{a\mu} - \phi^{\mu}\delta_{\psi}^{(0)}\alpha^{\nu (2)}_{a\mu}+ \nonumber \\
(\psi^{\mu}\partial^{b}\phi^{\zeta}-\phi^{\mu}\partial^{b}\psi^{\zeta})[f^{\sigma}_{\zeta \epsilon}d^{\epsilon \nu}_{\mu}\epsilon_{abc}A^{c}_{\sigma}+d^{\nu}_{\mu \epsilon}d^{\sigma \epsilon}_{\zeta}\epsilon_{abc}\epsilon^{c}_{de}\partial^{d}A^{e}_{\sigma}] = \partial_{a}\chi^{(1)\nu}
\end{align}
In (\ref{82}) we cannot set $\alpha^{\nu (2)}_{a\mu}$ and $\beta^{(2)b\nu}_{a\mu}$ to zero. This would mean that at higher orders, $\alpha$ and $\beta$ will also be non-zero in general. This implies that the transformation for the $A^{\mu}_a$ contains infinite number of terms, of arbitrarily higher order in $A$\footnote{It is possible, but unlikely, that this transformations truncates at some higher order.}

\begin{equation}\label{nontrivial}
\delta A^{\mu}_{a} = \partial_{a}\chi^{\mu} + f^{\mu}_{\lambda\sigma}A^{\sigma}_{a}\chi^{\lambda} + d^{\nu\mu}_{\lambda}\epsilon_{abc}\partial^b A^{c}_{\nu}\chi^{\lambda} +\mathcal{O}(A^2)\chi+\mathcal{O}(A^2)\partial\hspace{.05 mm}\chi
\end{equation}
\end{itemize}

\subsection{Analysis of \ref{73}-\ref{75}}\label{analysis}

Depending on whether we choose $d=0$ or not, we end up with two distinct cases, 
\begin{itemize}
\item[1] If we choose $d=0$,  then all the non linear terms in gauge transformation vanish and we end up with the standard non abelian gauge transformations, or in the special case with $f=0$, abelian gauge transformations.
\item[2] If $d\neq 0$  we will have the new non linear transformation \ref{nontrivial}, satisfying the equations (\ref{73}-\ref{75}).  
\begin{itemize}
\item[(2a)] Special case of this is when $f=0$. In particular for $k=1$, we have $f=0$. \textit{In a previous version of the paper, we had wrongly claimed that in this case, $d=0$. Our equations were correct, however we missed a non trivial $d\neq 0$ solution}. In fact $k=1$ is the simplest case with $d\neq 0$ solution of the equation \ref{74}. Its easy to see that the equation\ref{74} is trivially satisfied, for $k=1$. \textit{ In fact for this case, the analysis has been done long back by Heiderich and Unruh\cite{Heiderich:1990bb}. We had missed this reference previously. Our analysis reduces to theirs in the $k=1$ limit}. We can also have $k \neq 1$ and $f = 0$, then the d’s need only satisfy
equation \ref{74} In this case ofcourse there will be solutions, the simplest being
when we put all $d^{\mu\nu}_{\lambda} = \kappa$ for all$(\mu, \nu, \lambda)$.
\item [(2b)] More generally, we can ask whether there are solutions for $k\neq 1$ and $f \neq 0$.  Solutions also exist for this case, as we show below by explicitly analyzing the equations \ref{73}-\ref{75} for some low values of $k$- the number of species, eg: $k=2$ and $k=3$ and also $k=6$.
\end{itemize}
  
\end{itemize}

\begin{itemize}

\item \textbf{$k = 2$}

In this case, we have a unique non abelian non compact, non simple lie algebra, with two generators- $[X, Y] =Y$, with structure constants $f^{1}_{12}=1$ and rest vanishing\footnote{One place where this algebra appears is when we consider the set of translations and rotation in two dimensions. These are generated by $T_x =i\frac{\partial}{\partial x}$, $T_y =i\frac{\partial}{\partial y}$ and $R= i(x\frac{\partial}{\partial y} -y\frac{\partial}{\partial x})$. The corresponding algebra is $[R,T_x] = -iT_y,\; \; [R, T_y] = iT_x ,\; \; [T_x, T_y] =0$. Its then easy to check that $[R, T^{-}] = T^{-}$, with $T^{-} = T_x -iT_y$.}. 
While Yang Mills theories are based on semi-simple lie algebras, there has been some studies on gauge theories based on non simple lie algebras in the literature as well, see for eg:\cite{Tseytlin:1995yw}.

In this case there are 8 d's i.e $d^{11}_{1}$,$d^{12}_{1}$,$d^{21}_{1}$,$d^{22}_{1}$,$d^{11}_{2}$,$d^{12}_{2}$,$d^{21}_{2}$,$d^{22}_{2}$. The d's satisfy the two equations \ref{74} and \ref{75}. From \ref{75} we get the following four equations:
\begin{align}
d^{12}_{2} - d^{11}_{1} - d^{21}_{2} = 0 \\
d^{22}_{2} = 0 \\
d^{12}_{1} = 0 \\
d^{22}_{1} = 0 
\end{align}
Using the above equations we get the following four equations by solving \ref{74}
\begin{align}
d^{21}_{1}d^{12}_{2} = 0 \\
d^{21}_{2}(d^{11}_{1} - d^{11}_{2}) = 0 \\
d^{12}_{2}d^{11}_{1} = 0 \\
d^{21}_{1}d^{12}_{2} = 0
\end{align}
From the last four equations, we may either take $d^{21}_{1} = d^{21}_{2} = d^{11}_{1} = 0$ with non zero arbitrary $d^{11}_{2}$ or, $d^{12}_{2} = 0$ with arbitrary non vanishing $d^{11}_{1} = d^{11}_{2} = -d^{21}_{2}$. Therefore there are solutions with non zero $d$'s.

\item \textbf{$k = 3$}

In the above example the lie algebra was neither compact nor simple. We will now consider the example of  simple compact lie algebra- $SU(2)$. For $k = 3$ and  $SU(2)$ structure constants, we have analyzed the equations explicitly. In this case, the structure constant of the Lie algebra or $f^{\nu}_{\mu\lambda}$ will be $\epsilon_{\mu\nu\lambda}$ which satisfies the Jacobi identity (\ref{73}). Here each $\mu$,$\nu$,$\lambda$ runs from 1 to 3. Now using 27 equations of (\ref{75}) we get the following constraints on 27 d's:
\begin{align}\label{3.34}
d^{2}_{33} = d^{3}_{32} = -d^{2}_{22}, \; \text{and all the other d's = 0}
\end{align}
Now we will consider the identity (\ref{74}) where we have 27 equations with 27 variables in the present case. After lowering the indices this identity becomes \\
\begin{align}
d^{\nu}_{\mu \epsilon}d^{\epsilon}_{\zeta\sigma}-d^{\nu}_{\zeta \epsilon}d^{\epsilon}_{\mu\sigma} = 0
\end{align}
Consider the following equation for the specific choice of $\nu=2,\mu=2,\zeta=3,\sigma=3$ and we get,
\begin{align}
d^{2}_{22}d^{2}_{33}-d^{2}_{33}d^{3}_{23}=0 \nonumber \\
\implies d^{2}_{22} = 0 
\end{align}
Using \ref{3.34} we can thus set all the d's to zero. Therefore in $SU(2)$ case there are no solutions with non-zero $d$'s.
 
\item \textbf{$k = 6$}

Since we have analyzed the case of $SU(2)$, it is natural to ask whether the $d$'s vanish for the case of $SO(4)$. Since this is isomorphic to $SU(2)\times SU(2)$, the $f$'s in this case nicely split into two sets. $f^{\gamma}_{\alpha\beta}=\epsilon_{\alpha\beta\gamma}$ and $f^{c}_{ab}=\epsilon_{abc}$. $(\alpha,\beta,\gamma =1,2,3)$ and $(a,b,c =4,5,6)$. However, its not necessary that the $d$'s split as well. There could, in principle be $d$'s with mixed indices of the type- $d^{\alpha \beta}_{c}$, $d^{\alpha b}_{c}$, $d^{a \beta}_{c}$, $d^{a \beta}_{\gamma}$, $d^{\alpha b}_{\gamma}$ and $d^{ab}_{\gamma}$. If these vanish, then the analysis reduces to the $SU(2)$ case, for which we know all the $d$'s vanish. We will show that there indeed are non vanishing solutions to the mixed $d$'s, which satisfies the equations \ref{74} and\ref{75}. In particular, equation \ref{75} gives the following constrains on the mixed $d$'s.
\begin{eqnarray}\label{so4}
&& d^{\alpha \beta}_{c} = d^{ab}_{\gamma} =  d^{\alpha c}_{c}= d^{c \beta}_{c}= d^{\alpha b}_{\alpha} = d^{a \beta}_{\beta} =0, \nonumber \\
&& d^{a \beta}_{\gamma} = -d^{a \gamma}_{\beta}, \; d^{\beta a}_{\gamma} = -d^{\gamma a}_{\beta}, \; d^{a \beta}_{b} = -d^{b \beta}_{a}, \; d^{\beta a}_{b} = -d^{\beta b}_{a} \nonumber \\
&& d^{41}_{5} = d^{63}_{2},\; d^{14}_{5} = d^{36}_{2},\; d^{14}_{6} = d^{25}_{3},\; d^{41}_{6} =  d^{52}_{3},\; d^{51}_{6} =  d^{43}_{2},\; d^{15}_{6} =  d^{34}_{2}, \nonumber \\
 && d^{16}_{2} = d^{35}_{4},\; d^{61}_{2} = d^{53}_{4},\; d^{16}_{3} = d^{24}_{5},\; d^{61}_{3} =  d^{42}_{5},\; d^{43}_{6} =  d^{51}_{2},\; d^{34}_{6} =  d^{15}_{2}, \nonumber \\
 && d^{53}_{6} = d^{42}_{1},\; d^{35}_{6} = d^{24}_{1},\; d^{42}_{6} =  d^{53}_{1},\; d^{24}_{6} =  d^{35}_{1},\; d^{52}_{6} =  d^{41}_{3},\; d^{25}_{6} = d^{14}_{3} 
\end{eqnarray}
With these constraints, it can be shown that setting the $d$'s of the type $d^{a\alpha}_{\beta}$ non vanishing and arbitrary, while setting all the rest not related to these via eq \ref{so4} to zero solves all the equations \ref{74}. Therefore there are solutions with non vanishing $d$'s case, in this case. 

Thus we have aleast one explicit example with a $f$ corresponding to a semi-simple lie algebra and non zero $d$'s. 
\end{itemize} 

\subsection{Comments on Lagrangians invariant under new gauge transformations}\label{lag}

We saw in the previous sections that Walds analysis gives as a solution to the integrability condition\ref{int}, new nonlinear gauge transformation with $d \neq 0$ which does not seem to truncate at any finite order. The existance of such gauge transformation does not necessarily mean that there exist Lagrangians which are invariant under it. Infact results from a cohomological analysis imply that the only consistent ineracting theories of spin one fields must have usual non-abelian gauge invariance\cite{Henneaux:1997bm}. 

One way to addressing this question, is to start by writing the most general Lagrangian, and then constraining the coefficients of various terms appearing in the Lagrangian, by demanding that under the gauge transformations\ref{nontrivial} consistent with eqns\ref{73}-\ref{75}, the Lagrangian should change atmost by a total derivative. 

This method of finding nonlinear generalizations of gauge theories, by simultaneously deforming both the Lagrangian as well as the gauge transformation, and demanding the invariance of the resulting action under these deformed gauge transformations is well known in the literature. See\cite{Anco:2004em} for a nice discussion of this method in a very general context involving p-form field theories in d- dimensions. 

Using this method, in a nice paper \cite{Anco:1995wt}, Anco has found the most general nonlinear Lagrangian of spin one fields for the case of $k\geq 1$ but $f=0$. The nonlinear deformations he finds corresponds to gauge transformations which close on-shell, and thus do not satisfy the constraint\ref{74}, which arises from solving\ref{int} offshell. Thus, atleast for this case, the only gauge invariant lagrangians, subject to the constraint \ref{73}-\ref{75}, correspond to the usual abelian gauge theories, for which $d=0$.\footnote{It should be noted that Anco was not looking for nonlinear gauge transformaions which close offshell, so he doesnot have the analogue of the equations\ref{73}-\ref{75}.}

We need to carry out this procedure for the more general case  of $k\neq 1$, $f\neq 0$ and $d\neq 0$. In this section we set up this calculation.

Since the quadratic piece of the Lagrangian is fixed to be either Maxwell or Chern-simons, we start by parametrizing the most general cubic term of the Lagrangian $\mathcal{L}^{(3)}$ as follows.

\begin{align}\label{L3}
\mathcal{L}^{(3)} = g_{\mu\nu\lambda}^{(1)} \epsilon_{abc}A^{a\mu}A^{b\nu}A^{c\lambda} +g_{\mu\nu\lambda}^{(2)}\partial_{a}A^{\mu a}A^{\nu b}A^{\lambda}_{b} +  g^{(3)}_{\mu\nu\lambda}\partial_{a}A^{\mu b}\partial^{a}A^{\nu c}A^{\lambda d}\epsilon_{bcd} +  \nonumber \\ g^{(4)}_{\mu\nu\lambda}\partial_{a}A^{a\mu}\partial^{b}A^{c\nu}A^{d\lambda}\epsilon_{bcd}+ g^{(5)}_{\mu\nu\lambda}\partial^{b}A^{a\mu}\partial^{c}A^{d\nu}A^{\lambda}_{a}\epsilon_{bcd} + g^{(6)}_{\mu\nu\lambda}\partial_{a}A^{a\mu}\partial_{b}A^{b\nu}\partial_{c}A^{c\lambda}
\end{align}
Where we have resricted the Lagrangian to terms upto three derivatives.
The $g^{(i)}$'s are arbitrary coefficients which we need to fix by imposing the condition of gauge invariance. More precisely:
\begin{equation}\label{laginv}
\delta_{(1)}\mathcal{L}^{(2)} + \delta_{(0)}\mathcal{L}^{(3)}= \textrm{total derivative}.
\end{equation}
 Here $\delta_{(n)}\mathcal{L}^{(m)} = \frac{\partial \mathcal{L}^{(m)}}{\partial A^{\mu}_{a}} \delta^{(n)}A^{\mu}_{a} + \frac{\partial \mathcal{L}^{(m)}}{\partial(\partial_{e} A^{\mu}_{a})} \delta^{(n)}(\partial_{e} A^{\mu}_{a})$ and $\delta_{(0)} A^{\mu}_{a}=\partial_{a}\chi^{\mu} $ and $\delta_{(1)} A^{\mu}_{a}= f^{\mu}_{\lambda\sigma}A^{\sigma}_{a}\chi^{\lambda} + d^{\mu}_{\lambda\sigma}\epsilon_{abc}\partial^{b}A^{c\sigma}\chi^{\lambda}$.
Using equations \ref{73}-\ref{75} and the antisymmetric property of $f^{\mu}_{\nu\lambda}$, we need to find constraints on the coefficients $g^{(i)}_{\mu\nu\lambda}\; (i=1,2..,5)$, such that equation\ref{laginv} is satisfied. One has to do this order by order at all orders. For instance, at the next order, we would demand that:
 \begin{equation}\label{L4}
 \delta_{(2)}\mathcal{L}^{(2)} + \delta_{(1)}\mathcal{L}^{(3)} + \delta_{(0)}\mathcal{L}^{(4)} = \textrm{total derivative}.
 \end{equation}
 
In the next section, we solve \ref{laginv}. We have used the Maxwell form for $\mathcal{L}^{(2)}$ in the analysis below. For notational simplicity, we also use all lower indices on $d_{\mu\nu\lambda}$. Comparing with our previous notation, the $d_{\lambda\mu\nu}\equiv d^{\nu\lambda}_{\mu}$
 \subsection{Determining $\mathcal{L}^{(3)}$ from \ref{laginv}} \label{A}
Before proceeding with the calculation, we first see what are the constraints on $g^{(i)}_{\mu\nu\lambda}$'s from the definition in the Lagrangian \ref{L3}. From the general structure of this Lagrangian it is clear that $g^{(1)}_{\mu\nu\lambda}$ is completely antisymmetric in all indices, $g^{(6)}_{\mu\nu\lambda}$ is completely symmetric while  $g^{(2)}_{\mu\nu\lambda} = g^{(2)}_{\mu\lambda\nu}, \; g^{(3)}_{\mu\nu\lambda} = -g^{(3)}_{\nu\mu\lambda}$

Similarly, upto total derivative terms, the $g^{(5)}$ term can be rewritten as:
\begin{align}
g^{(5)}_{\mu\nu\lambda}\partial^{b}A^{a\mu}\partial^{c}A^{d\nu}A^{\lambda}_{a}\epsilon_{bcd} = \text{total derivative term} - g^{(5)}_{\mu\nu\lambda}\partial^{b}A^{a\mu}A^{d\nu} \partial^{c} A^{\lambda}_{a}\epsilon_{bcd} \nonumber 
\end{align}
Therefore without loss of generality we can choose $g^{(5)}_{\mu\nu\lambda}$ to be antisymmetric in $\mu$ and $\lambda$.

 To solve equation \ref{laginv}, let us first consider the term $\delta_{(0)}\mathcal{L}^{(3)}$.
\begin{align}
\delta_{(0)}\mathcal{L}^{(3)} = (g^{(1)}_{\sigma\nu\lambda}\epsilon_{fbc}A^{b\nu}A^{c\lambda}+ g^{(1)}_{\mu\sigma\lambda}\epsilon_{bfc}A^{b\mu}A^{c\lambda} + g^{(1)}_{\mu\nu\sigma}\epsilon_{abf}A^{a\mu}A^{b\nu})\partial^{f}\chi^{\sigma} +\nonumber \\
(g^{(2)}_{\mu\sigma\lambda}\partial_{a}A^{a\mu}A^{f\lambda}+ g^{(2)}_{\mu\nu\sigma}\partial_{a}A^{a\mu}A^{f\nu})\partial_{f}\chi^{\sigma} +g^{(2)}_{\sigma\nu\lambda}A^{b\mu}A^{\lambda}_{b}\Box \chi^{\sigma} + \nonumber \\
(g^{(3)}_{\mu\nu\sigma}\partial_{a}A^{b\mu}\partial^{a} A^{c\nu}\epsilon_{bcf})\partial^{f}\chi^{\sigma} +(g^{(3)}_{\sigma\nu\lambda}\partial^{e}A^{c\nu} A^{d\lambda}\epsilon_{fcd} + g^{(3)}_{\mu\sigma\lambda}\partial^{e}A^{b\mu} A^{d\lambda}\epsilon_{bfd})\partial_{e}\partial_{f}\chi^{\sigma} + \nonumber \\
(g^{(4)}_{\sigma\nu\lambda}\partial^{b}A^{c\nu}A^{d\lambda}\epsilon_{bcd})\Box \chi^{\sigma} + (g^{(4)}_{\mu\nu\sigma}\partial_{a}A^{a\mu}\partial^{b} A^{c\nu}\epsilon_{bcf})\partial^{f}\chi^{\sigma} + \nonumber \\
(g^{(5)}_{\mu\nu\sigma}\partial^{b}A^{f\mu}\partial^{c} A^{d\nu}\epsilon_{bcd})\partial_{f}\chi^{\sigma} +(g^{(5)}_{\sigma\nu\lambda}\partial^{c}A^{d\nu}A^{f\lambda} \epsilon_{ecd})\partial^{e}\partial_{f}\chi^{\sigma} + \nonumber \\
+(g^{(6)}_{\sigma\nu\lambda}\partial_{b}A^{b\nu}\partial_{c}A^{c\lambda}+ g^{(6)}_{\mu\sigma\lambda}\partial_{b}A^{b\mu}\partial_{c}A^{c\lambda} + g^{(6)}_{\mu\nu\sigma}\partial_{b}A^{b\mu}\partial_{c}A^{c\nu})\Box \chi^{\sigma}
\end{align}
Similarly from the other variation of the Lagrangian $\mathcal{L}^{(2)}$ we get,
\begin{align}
\delta_{(1)}\mathcal{L}^{(2)} = -(f_{\mu\lambda\sigma}A^{\sigma}_{a}\chi^{\lambda} + d_{\mu\lambda\sigma}\epsilon_{abc}\partial^{b}A^{c\sigma}\chi^{\lambda})\partial_{e}F^{ea\mu}
\end{align}
Here $f$ satisfies the antisymmetry property $f_{\mu\nu\lambda} =-f_{\mu\lambda\nu}$. Adding these two parts and demanding that it adds up to a total derivative, we will end up with the following constraints,
\begin{align}
g_{\mu\nu\lambda}^{(1)} = g_{\mu\nu\lambda}^{(6)} = g_{\mu(\nu\lambda)}^{(2)} = g^{(5)}_{\mu\nu\lambda} = 0 \\
g_{\mu[\nu\lambda]}^{(2)} = f_{\mu\nu\lambda} \\
g_{\mu\nu\lambda}^{(3)} = \frac{1}{2}g_{\mu\nu\lambda}^{(4)} = \frac{1}{2}d_{\mu\lambda\nu} \\
d_{\mu\lambda\nu} = - d_{\nu\lambda\mu} \\
\end{align}
Therefore  we can rewrite all the undetermined $g^{(i)}$'s  in terms of f and d. The most non trivial result is the antisymmetry in d i.e $d_{\mu\lambda\nu} = - d_{\nu\lambda\mu}$. In the notation used in the previous section, this is the condition that $d^{bc}_{a} = -d^{cb}_{a}$. 

For the case of $SO(4)$, that we had analyzed, imposing this property, kills all the $d$'s. Thus atleast in this limited context we see how demanding the existance of a Lagrangian, puts $d=0$.
In this way we get the following form of $\mathcal{L}^{(3)}$
\begin{align}
\mathcal{L}^{(3)} = -f_{\mu\nu\lambda}\partial_{a}A^{\nu b}A^{\mu a}A^{\lambda}_{b} +  \frac{1}{2}d_{\mu\lambda\nu}\partial_{a}A^{\mu b}\partial^{a}A^{\nu c}A^{\lambda d}\epsilon_{bcd} + d_{\mu\lambda\nu}\partial_{a}A^{a\mu}\partial^{b}A^{c\nu}A^{d\lambda}\epsilon_{bcd}
\end{align} 

Note that at this order, we do not need to use the equations (\ref{73}-\ref{75}). These equations
will play a role to determine $\mathcal{L}^{(4)}$ via \ref{L4}. But for this we need to know $\delta_{(2)}A^{a\mu}$, but, in
this note, we have only considered the form of non linear gauge transformation upto first
order in $A^{a\mu}$, and so are unable to solve \ref{L4}. These might provide further constraints
on the $d’$s. It is possible that the constraints on $d$ from this analysis coupled with the
equations (\ref{73}-\ref{75}), are enough to put all $d=0$. For instance, already we see that the
non zero $d$ solution both for the $SO(4)$ case, as well as the $k = 2$ case, that we presented
in the previous section, is killed by the antisymmetry property of $d$. This is so because,
the antisymmetry condition equates the non zero $d$’s with the ones which vanish. This
suggests that might happen more generally. It would be nice to check this explicitly.

\section{Theories with non-diagonal kinetic terms- BF theory} \label{BF}
Finally, we end by noting that we can generalize the above discussion to include theories with non-diagonal kinetic term. 
\begin{equation}
\mathcal{L}^{(2)} = \epsilon^{abc}h_{\mu \nu}A^{\mu}_{a}\partial_{b}A^{\nu}_{c} \;
\textrm{or,} \; \; \mathcal{L}^{(2)} = h_{\mu\nu}\mathcal{F}^{\mu}_{ab}\mathcal{F}^{\nu ab}
 \end{equation}
Or a linear combination of both \footnote{$h_{\mu\nu}$ must have non-vanishing eigenvalues for this to reduce to Wald's case after diagonalization, else some of the fields will have no kinetic term}. In all such cases, the divergence condition $\partial_{a}\mathcal{F}^{(1)\mu a} =0$ is still satisfied, so that Wald's method as used in this paper, will go through without any significant modification. We expect therefore that the results derived in last section will be valid even in this case. The extention to non-diagonal case helps us to extend the analysis to BF theories. 

To get to BF theory, we need to start with even number of species of gauge fields $\mu=1,...N, N+1,...2N$ and label the first $N$ fields as $A^{\mu}$ and the last set of $N$ fields as $B^{\mu}$ and take the Chern-simons kinetic term with  $h_{\mu\nu}$ of the form:
\[ h_{\mu\nu} = \left( \begin{array}{cc}
0_{N\times N} & 1_{N\times N}  \\
1_{N\times N} & 0_{N\times N}  \end{array} \right)\]

Choosing the solution, $f^{\mu}_{\nu\lambda} =f^{\mu +N}_{\nu(\lambda +N)}$ ($\mu, \nu\, \lambda =1,...N$) non vanishing and all other $f=0$, that are not related by symmetry\footnote{It is easy to see that the remaining non vanishing $f$'s satisfy the Jacobi identity}, as well as $d=0$, gives us the gauge invariance of the BF theory\footnote{One place where the gauge symmetries are explicitly given is \cite{Mukhi:2011jp}.}.
\section{Summary and discussion of results} \label{sum}
To summarize, we have applied Wald's analysis to the case of spin-one fields in three dimensions. The technical difference from Wald's case is the role played by the three dimensional antisymmetric tensor $\epsilon^{abc}$. In our analysis we have studied the gauge invariance of all possible spin one fields in three dimension i.e Maxwell,Chern-Simons and BF field. We get known infinitesimal gauge invariances in all these theories as a special case $d=0$. 

We show that the  $\epsilon^{abc}$ term changes the analysis. In particular, the gauge transformation does not get truncated to linear order in $A$, and we do not in general get non-abelian lie-algebra valued gauge transformations by applying Wald's procedure. However one consistent solution is obtained by putting the coefficient of the  $\epsilon^{abc}$ term to zero, in which case, the analysis reduces to that of Wald. A special case of this, corresponding to a single spin-one field, has already been analyzed in the literature in\cite{Heiderich:1990bb}. Our analysis reduces to theirs in this special case.

One can ask, whether introducing a similar term in four dimensions- ie:- the four dimensional antisymmetric $\epsilon^{abcd}$, changes Wald's analysis or conclusions. We can see that this is not the case. The point is that, its easy to see that we cannot construct any coefficients of first order $\alpha_a^{(1)}$ and $\beta^{b(1)}_a$ using $\epsilon_{abcd}$. Therefore by using zeroth and first order part of integrability condition we can always make higher order $\alpha$ and $\beta$ to be zero without any effect of $\epsilon_{abcd}$, thus reducing the analysis to the one by Wald. 

We have also explicitly solved for $d$'s in three cases corresponding to  species number $k=2$, $k=3$ and $k=6$. For $k=2$ and $f^1_{12}\neq 0$, which corresponds to a non compact non simple lie algebra, there are solutions with non zero $d$'s. For $k=3$ and SU(2) structure constants, we explicitly showed that the coefficient of the  $\epsilon^{abc}$ term vanishes, so that we get back the usual non-abelian gauge invariance. However for $k=6$ and $SO(4)$ lie algebra, there are solutions again with non zero $d$'s.

Following \cite{Anco:1995wt}, we try to analyze the constraints coming from the existance of a Lagranian invariant under these new nonlinear gauge transformations. At the lowest order, this puts a constraint on the $d$'s. We see that this constraint is enough to kill the explicit
non zero $d$ solutions that we presented in a previous section, for the case of $SO(4)$ as well
as for $k = 2$ example. To carry out this analysis beyond the lowest order, we need to work
out the form of the non linear gauge transformations to the next order. It is very likely
that the constraints at this order coupled with the constraints coming from \ref{73}-\ref{75},
will imply the vanishing of the $d$’s. It would be nice to check this out explicitly.

\vspace*{1ex}
\noindent{\bf Acknowledgment:} 

We would like to thank collectively Samrat Bhowmick, Dileep Jatkar, Parthasarathi Majumdar, Pushan Majumdar and Koushik Ray for useful discussions. The research work of Suchetan Das is supported by a fellowship from CSIR. Finally, we would like to thank the anonymous referee of JHEP, whose comments and suggestions, have been very helpful to us.

\end{document}